\documentclass{article}
\usepackage{graphicx,color,epsfig,subfig,dcolumn,bm,mathrsfs,amsmath}

\newcommand{\change}[1]{\textcolor{black}{#1}}

\begin{document}

\title{Structure formation in soft matter solutions induced by solvent evaporation}

\begin{center}
\textbf{Structure Formation in Soft Matter \change{Solutions} Induced by Solvent Evaporation}\\
\vspace{0.5cm}

Jiajia Zhou $^{1,3,\dagger}$, Xingkun Man $^{2,3,\dagger}$, Ying Jiang $^{1,3,\dagger}$, and Masao Doi $^{3,4,*}$\\
\vspace{0.5cm}

$^1$ Key Laboratory of Bio-inspired Smart Interfacial Science and Technology of Ministry of Education, School of Chemistry, Beihang University, Beijing 100191, China\\
$^2$ School of Physics and Nuclear Energy Engineering, Beihang University, Beijing 100191, China\\
$^3$ Center of Soft Matter Physics and Its Applications, Beihang University, Beijing 100191, China\\
$^4$ Beijing Advanced Innovation Center for Biomedical Engineering, Beihang University, Beijing 100191, China\\
\vspace{0.2cm}
$*$ corresponding author: masao.doi@buaa.edu.cn\\
${\dagger}$ equal contribution\\
\end{center}


\begin{abstract}
Solvent evaporation in soft matter solutions (solutions of colloidal particles, polymers and their mixtures) is an important process in material making and in printing and coating industries. 
The solvent evaporation process determines the structure of the materials, and strongly affects their performance.  
Solvent evaporation involves many physico-chemical processes: flow, diffusion, \change{crystallization}, gelation, glass transition, etc., and is quite complex.
In this article, we report recent progresses in this important process, with a special focus on \change{theoretical} and simulation studies.  
\end{abstract}

\section{Introduction}

Solvent evaporation is a phenomenon we see in our daily life. 
One can observe this phenomenon when watching the drying paint \cite{vanderKooij2015} or the ring-like deposit of coffee stain. \cite{Deegan1997, ManXingkun2016}
It is a process widely used in many industries: chemical, material, food, and pharmaceutical industries. 
It is also important in technological frontiers, such as large-scale manufacture of device by printing and coating, and development of advanced materials. 

Solvent evaporation is generally considered to be a process of diffusion, 
the transport of nonvolatile solvents driven by chemical potential gradient created by the evaporation. 
However, the process is intrinsically non-equilibrium, and the simplicity in the practical procedure sometimes disguises the complexity of the underlying mechanisms. 
The drying phenomenon actually involves many physico-chemical processes such as flow, diffusion, phase separation, crystallization, gelation, glass transition, etc., and the interplay between different processes determines the final structure of the dried materials.

In this Research News, we report recent advances of self-assembled structures induced by uniform evaporation, examples including drying in a thin-film geometry or evaporating a spherical droplet on a superhydrophobic surface. \cite{Feng2002} 
In this situation, the structure formation takes place in the direction of the evaporation. 
This excludes the case where lateral flow is induced by non-uniform drying.
The lateral flow is important in explaining the coffee-ring effect, where excellent reviews exist in literature. \cite{HanWei2012, Larson2014}
After introducing the \change{general procedure in experiments}, we explain the physical mechanisms responsible for the structure formation in solvent evaporation. 
We then discuss recent advances based on single- and binary-component solutions, with a special emphasis on the \change{theoretical} approaches.

\section{Drying of single-component solutions}

\change{A general experiment starts with spin-coating a thin film solution on a substrate, then the sample is left to dry.
In the first step, it is important that the spin-cast film is uniform. \cite{Bornside1989, Fowler2016} 
In the second step, the evaporation flux, i.e., the mass removed from the liquid phase to the vapor phase in unit time from unit surface area, needs to be controlled precisely. 
Even for the pure solvent evaporation, the process is complex and involves the diffusion of solvent molecules in both the liquid and vapor phases, and external flow in the vapor phase to prevent saturation.
The evaporation flux can be computed using the Hertz-Knudsen equation which is based on the thermodynamics variables, \cite{Eames1997, Marek2001, Holyst2015} or solving the diffusion equation in the vapor phase using suitable boundary conditions. \cite{Lebedev, Picknett1977, HuHua2002}}

\begin{figure}[htbp]
  \centering
  \includegraphics[width=1.0\textwidth]{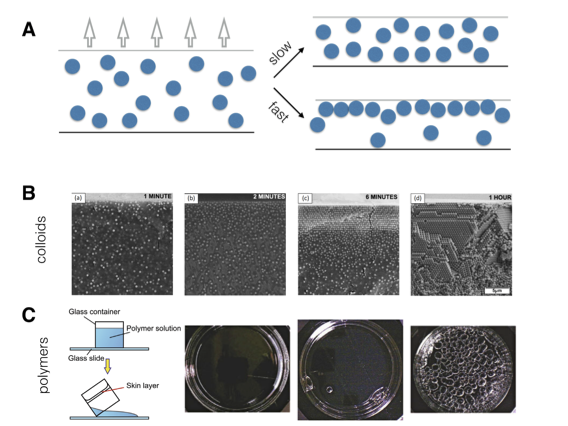}
  \caption{{\bf A}. The film structure after evaporation is determined by the evaporation rate. {\bf B}. Cryo-SEM images showing a consolidation front in a evaporating film composed of polystyrene particles in water. Reproduced with permission. \cite{Ma2005} Copyright 2005 Elsevier. {\bf C}. Top views of a evaporating PVAc/aceton solution. Bubbles can be seen in the last picture. Reproduced with permission. \cite{Arai2012} Copyright 2012 Springer.}
  \label{fig:single}
\end{figure}

The solvent evaporation drives the free surface to recede towards the substrate, resulting in an accumulation of the solutes at the surface.
Two competing processes determine the spatial distribution of the solutes.
One is the Brownian motion, which is characterized by the diffusion constant $D$ of the solutes.
The other is the evaporation, characterized by the rate $v_{\rm ev}$ at which the free surface recedes.  
Two time scales are related to these two process: the diffusion time $\tau_{\rm D} = \ell^2/D$, where $\ell$ is a characteristic length, and the evaporation time $\tau_{\rm ev}=\ell/v_{\rm ev}$. 
To quantify the relative importance between the diffusion and the evaporation, one can define a dimensionless number called film formation Peclet number \cite{Routh2013}
\begin{equation}
  {\rm Pe} = \frac{\tau_{\rm D}}{\tau_{\rm ev}}
           = \frac{ v_{\rm ev} \ell }{D} . \nonumber
\end{equation}
If ${\rm Pe}<1$, the diffusion takes place quickly; the concentration gradient created by evaporation is quickly flattened by diffusion, leading to a uniform concentration profile. 
If ${\rm Pe}>1$, the evaporation dominates; the concentration gradient increases with time, and the solutes accumulate at the solution/air interface. 
These two different scenarios are shown schematically in Figure \ref{fig:single}A. 
The time evolution of the film structures are shown for the evaporating colloidal dispersion \cite{Ma2005} (Figure \ref{fig:single}B) and polymer solution \cite{Arai2012} (Figure \ref{fig:single}C), respectively.


The final structure of the dried material depend on the evaporation rate. 
The general consensus is when the evaporate is slow, the solutes have ample time to adjust and self-organize, resulting in a more ordered structure. 
Otherwise, disordered structure forms. 
This provides an external means to control the final structure through the evaporation rate, which can be modified during the experiment. 

Industrial applications prefer fast evaporation because of the short production time. 
However, there are several new phenomena emerge at high evaporation rate.  
Fast evaporation increases the solute concentration at the surface, and the viscosity of this part increases.
When the concentration exceeds a certain critical value, the viscosity diverges, and the solutions at the surface becomes gel-like, which is called skin. \cite{Routh2013, Arai2012, Kajiya2006}
For applications, skin is usually considered to be avoided since it causes various unfavorable effects on the dried material, such as surface roughening \cite{Routh2013}, buckling and dimpling \cite{Bornside1989, deGennes2002, Pauchard2003, Tsapis2005, Boulogne2013a}, cavity formation \cite{Arai2012, Sadek2013, MengFanlong2014, MengFanlong2016a}
On the other hand, skins are utilized in some situations. 
For example, in spray drying, small droplets ejected from a nozzle quickly dry up and create a hollow or buckled particles due to skin formation, which is helpful in food industries. \cite{Masters}

Skin formation is a process that material changes from viscoelastic fluid to viscoelastic solid. 
There are many fundamental questions unanswered, such as when the skin formation starts, how skin grows in time, and how it affects the drying dynamics.  
Therefore, it is a challenge for both experimentalists and theorists. 
Measuring the change of rheological properties on a scale less than 1 $\mu$m is difficult in experiments.
Constructing a continuum mechanical theory for materials which transform from fluid to solid is a challenge for theorists.  
Okuzono et al. \cite{Okuzono2006} developed a simple theory to discuss the growth of the skin layer. 
They assumed that skin appears when solute concentration exceeds certain critical concentration and discussed the condition for the formation of skin layers and the growth dynamics of skin layer.  
A more detailed calculation was conducted based on a Lagrangian scheme. \cite{LuoLing2016}

As the evaporation proceeds, the skin layer is compressed, and the elastic energy of deformation increases. 
This causes various mechanical instabilities in the solution. 
A typical mechanical instability is buckling.
The former has been analyzed by Bornside et al. \cite{Bornside1989} and 
de Gennes \cite{deGennes2002}, and discussed intensively in the literature \cite{Lintingre2016}.
Another type of instability is cavitation.
In this case, the elastic energy is release by the appearance of gas bubbles inside the solution (Figure \ref{fig:single}C). 
Meng et al. developed a theory to model the cavitation process in the drying droplet. \cite{MengFanlong2014, MengFanlong2016a}
This theory showed that the shear modulus of the skin layer is essential for cavitation.

\change{
For colloidal dispersion, when the concentration increases further near the free surface, colloidal particles can form crystal. 
The quality of the crystal formation is found to be better at slow evaporation rate, while defects and grain boundaries start to appear at high evaporation rate. \cite{ChengShengfeng2013}. 
Novel method has been implemented to identify the crystalline structures during the evaporation. \cite{Reinhart2017}
}

\begin{figure}[htbp]
  \centering
  \includegraphics[width=1.0\textwidth]{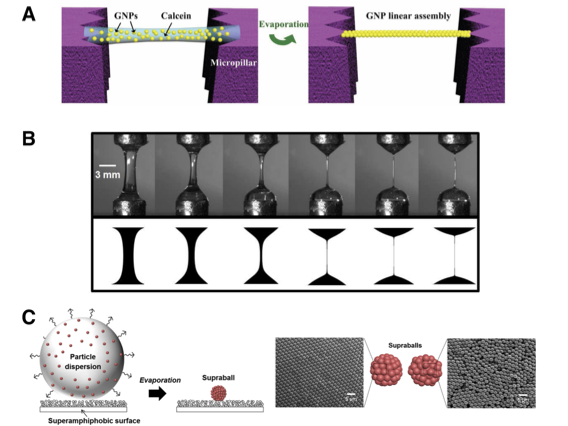}
  \caption{{\bf A}. Evaporation of a colloidal solution filament in between two micropillars. Reproduced with permission. \cite{SuBin2013} Copyright 2013 Wiley. {\bf B}. Evaporation of a polymer solution filament in extensional rheometer. Comparison between the experiment and numerical simulation are shown. Reproduced with permission. \cite{msc_Crest} Copyright 2009 MIT. {\bf C}. Evaporation of a spherical droplet. Supraballs of different ordering are produced under different pH values. Reproduced with permission. \cite{Sekido2017} Copyright 2017 ACS.}
  \label{fig:geo}
\end{figure}

The evaporation can also take place on a curved surface, as the solution takes a cylindrical or spherical geometry. 
Su et al. have developed a method to produce nanowires composed of polymers or nanoparticles. \cite{SuBin2012, SuBin2013}
Liquid bridges are formed in between micropillars. 
The cylindrical filament are subsequently dried and uniform structures emerge (Figure \ref{fig:geo}A). 
In the case of polymer solution, the initial polymer concentration is a vital controlling parameter; either too small or too large concentration causes the nanowires to break.  
A more controlled experiment was performed in extensional rheometer by Crest et al. \cite{msc_Crest} 
In this case, the competition between elasticity, evaporation, and capillarity needs to be carefully tuned (Figure \ref{fig:geo}B).
A solution droplet keeps a spherical shape on a superhydrophobic surface. 
When the spherical solution is dried \cite{Wooh2015, Sekido2017}, supraballs of different ordering and mechanical properties can be produced (Figure \ref{fig:geo}C).


\section{Drying of two-component solutions}

The one-component solution discussed in previous section is an ideal case. 
In \change{practical applications}, the solutions contain particles of different sizes or of different properties, surfactants, polymers, etc.
Each component contributes different functionality to the material.
To realize specific functionality, the arrangement of various species in the dried film has to be controlled precisely.
One desired case is the uniform distribution, as in nanocomposite, the nanoparticles need to be dispersed uniformly in the polymer matrix to enhance the mechanical property. \cite{Kumar2017} 
Another case is the layered structure, which is traditionally produced by layer-by-layer assembly. 
Evaporation can create multilayer coating during a single drying step, therefore it is highly desirable from industrial point of view. \cite{Routh2013}

\begin{figure}[htbp]
  \centering
  \includegraphics[width=1.0\textwidth]{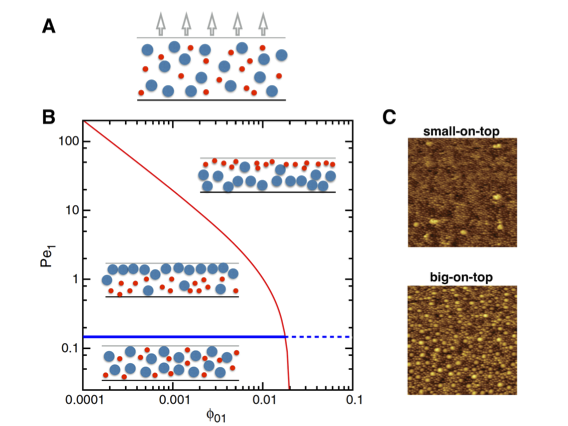}
  \caption{{\bf A}. Evaporation in a colloidal solution composed of big and small particles. {\bf B}. State diagram of final structure based on the initial concentration of small particles and the Peclet number. Three structures are predicted: small-on-top, big-on-top, and homogeneous structure. Reproduced with permission. \cite{Zhou2017} Copyright 2017 APS. {\bf C}. Experimental observations of small-on-top and big-on-top structures. Reproduced with permission. \cite{Trueman2012a} Copyright 2012 ACS.}
  \label{fig:binary}
\end{figure}

Consider a colloidal solution consisting of two kinds of particles, big and small (see Fig.~\ref{fig:binary}A).
In the solution, they mix each other uniformly.   
As solvent evaporates, the concentration at the surface becomes larger than in the bulk, and concentration gradient is created.  
This effect eventually creates a composition gradient in the dried film.
\change{Here we have two Peclet numbers, $\mathrm{Pe}_{1}$ for small colloids and $\mathrm{Pe}_{2}$ for big colloids.
For hard spheres, the Peclet numbers are related by the size ratio, $\mathrm{Pe}_{1}/\mathrm{Pe}_{2} = R_{1}/R_{2}$.}
When $\mathrm{Pe}_2 >1$ and $\mathrm{Pe}_1 < 1$, big particles are expected to remain at the surface and small particles can quickly diffuse away, giving the big-on-top stratification.  
Indeed, Trueman et al. \cite{Trueman2012, Trueman2012a} reported that the big-on-top stratification is observed in many colloidal systems. 
However, they also reported that when the size ratio becomes large
(roughly larger than 3), the inverted structure, small-on-top stratification is observed.  
They conjectured that this is due to the interaction between particles. \cite{Trueman2012} 
On the other hand, Fortini et al. \cite{Fortini2016} conducted a computer simulation and reported that the small-on-top structure appears for hard sphere systems \change{when both Peclet numbers are large ($\mathrm{Pe}_1 > 1$, $\mathrm{Pe}_2 > 1$) and the size ratio is large (larger than 7)}. 
Howard et al. \cite{Howard2017} performed a through examination of the parameter space on a similar system.


Zhou et al. recently proposed a simple explanation for the small-on-top structure. \cite{Zhou2017}
They assumed a hard sphere model, and accounted for the particle-interaction by the virial expansion for the free energy.  
Based on the standard diffusion model, \cite{Routh2004} they derived time evolution equations for particle concentrations, and showed that the small-on-top structure indeed appears for mixed suspensions with large size ratio.  
The key point is that the effect of interaction on the particles is not symmetric: the force exerted on big particles by small particles is much pronounced than the force exerted on small particles by big particles.  
Hence, if the concentration gradient is created at the surface, big particles feel larger forces than small particles and are pushed away from the surface.
A state diagram was constructed to indicate which type of stratification is created when evaporation rate and particle concentration are varied (Figure \ref{fig:binary}B). 
Their results are in quantitative agreement with the results of simulation and 
experiments \cite{Trueman2012a, Fortini2016}. 

\begin{figure}[htbp]
  \centering
  \includegraphics[width=1.0\textwidth]{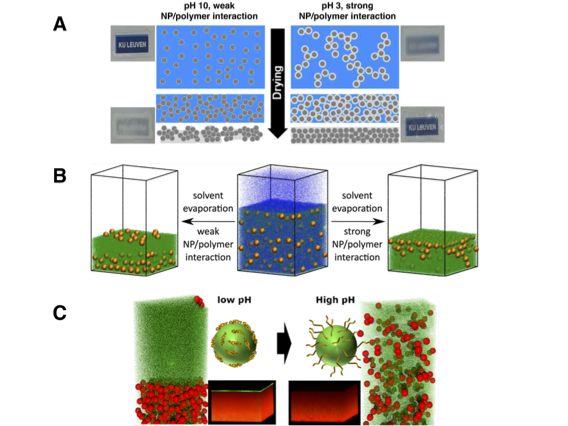}
  \caption{{\bf A}. Structure change by tuning the pH value, which changes the interaction between polymers and nanoparticles. Reproduced with permission. \cite{Kim2016b} Copyright 2016 ACS. {\bf B}. Structure change by tuning the attraction between polymers and nanoparticles. Reproduced with permission. \cite{ChengShengfeng2016} Copyright 2016 ACS. {\bf C}. Structure change by tuning the self-interaction of small particles. Reproduced with permission. \cite{Martin-Fabiani2016} Copyright 2016 ACS.}
  \label{fig:switch}
\end{figure}

These studies have highlighted the importance of particle interaction.
Another interesting system is the mixture of polymers and nanoparticles, in which their interaction can be varied via the pH. 
At low evaporation rate, Kim et al. \cite{Kim2016b} reported weak particle-polymer attraction produce dense aggregates of bare particles in the dried film, while strong particle-polymer attraction results in more homogeneous distribution (Figure \ref{fig:switch}A). 
At high evaporation rate, Cheng et al. \cite{ChengShengfeng2016} performed large-scale molecular dynamics simulations. 
Their results indicates that stratified structure can be produced by evaporation: poor particle-polymer affinity leads to polymer-on-top structure, while good particle-polymer affinity results in the opposite particle-on-top structure (Figure \ref{fig:switch}B). 
At fast evaporation, the skin layer is present and may be the reason for stratification. 


Interaction between particles of the same type can also be modified by colloidal surface property. 
Mart{\'i}n-Fabiani et al. \cite{Martin-Fabiani2016} synthesized nanoparticles with a hairy shell composed of hydrophilic PMAA chains. 
\change{By increasing the pH, the polymer chains change from a collapse state to extended state, effectively increase the particle size.}
When these hairy nanoparticles are mixed with large particles, the dried structure can be switched from homogeneous to layered by adjusting the pH value (Figure \ref{fig:switch}C).
On the \change{theoretical} side, Atmuri et al. developed a diffusive model which explicitly took into account the self-interaction between particles of the same type, but the cross-interaction is absent. \cite{Atmuri2012}

\section{Outlook}

Understanding the solvent evaporation process, especially how to control the structural change taking place during the evaporation is an important subject in many applications.
There are two advantages for solution-based method in comparison to methods without solvents:
(1) The solutes are kinetically mobile in the solution, therefore the dynamics is fast and fabrication time can be reduced. 
(2) The final structure can be controlled by external means such as evaporation rate and solution pH value.
If the layered structure is preferred, it is possible to create multi-layered coatings in one simple fabrication step. 
Here we have highlighted recent development in structure control through the evaporation rate and the solute interactions, with a special emphasis on the synergy between theory, simulation, and experiment. 
By combining different research approaches, we can have a better understanding of the governing principles of structure formation during evaporation and aid the development of advanced materials.

\section{Acknowledgments}
This work was supported by the National Natural Science Foundation of China (NSFC) through the Grant No. 21434001, 21404003, 21504004, 21574006, and 21622401, and the joint NSFC-ISF Research Program, jointly funded by the NSFC and the Israel Science Foundation (ISF) under grant No. 51561145002 (855/15).
M. D. acknowledges the financial support of the Chinese Central Government in the Thousand Talents Program.


\bibliographystyle{unsrt}
\bibliography{evaporation}

\begin{thebibliography}{10}

\bibitem{vanderKooij2015}
Hanne~M. van~der Kooij and Joris Sprakel.
\newblock Watching paint dry: more exciting than it seems.
\newblock {\em Soft Matter}, 11:6353--6359, 2015.

\bibitem{Deegan1997}
Robert~D. Deegan, Olgica Bakajin, Todd~F. Dupont, Greb Huber, Sidney~R. Nagel,
  and Thomas~A. Witten.
\newblock Capillary flow as the cause of ring stains from dried liquid drops.
\newblock {\em Nature}, 389:827, 1997.

\bibitem{ManXingkun2016}
Xingkun Man and Masao Doi.
\newblock Ring to mountain transition in deposition pattern of drying droplets.
\newblock {\em Phys. Rev. Lett.}, 116:066101, 2016.

\bibitem{Feng2002}
L.~Feng, S.~Li, Y.~Li, H.~Li, L.~Zhang, J.~Zhai, Y.~Song, B.~Liu, L.~Jiang, and
  D.~Zhu.
\newblock Super-hydrophobic surfaces: From natural to artificial.
\newblock {\em Adv. Mater.}, 14:1857, 2002.

\bibitem{HanWei2012}
Wei Han and Zhiqun Lin.
\newblock Learning from ``coffee rings'': Ordered structures enabled by
  controlled evaporative self-assembly.
\newblock {\em Angew. Chem. Int. Ed.}, 51:1534, 2012.

\bibitem{Larson2014}
Ronald~G. Larson.
\newblock Transport and deposition patterns in drying sessile droplets.
\newblock {\em {AlChE} Journal}, 60:1538, 2014.

\bibitem{Bornside1989}
D.~E. Bornside, C.~W. Macosko, and L.~E. Scriven.
\newblock Spin coating: One-dimensional model.
\newblock {\em J. Appl. Phys.}, 66:5185--5193, 1989.

\bibitem{Fowler2016}
Paul~D. Fowler, C{\'{e}}line Ruscher, Joshua~D. McGraw, James~A. Forrest, and
  Kari Dalnoki-Veress.
\newblock Controlling marangoni-induced instabilities in spin-cast polymer
  films: How to prepare uniform films.
\newblock {\em Eur. Phys. J. E}, 39(9):90, 2016.

\bibitem{Eames1997}
I.W. Eames, N.J. Marr, and H.~Sabir.
\newblock The evaporation coefficient of water: a review.
\newblock {\em Int. J. Heat Mass Transfer}, 40(12):2963--2973, aug 1997.

\bibitem{Marek2001}
R.~Marek and J.~Straub.
\newblock Analysis of the evaporation coefficient and the condensation
  coefficient of water.
\newblock {\em Int. J. Heat Mass Transfer}, 44(1):39--53, jan 2001.

\bibitem{Holyst2015}
Robert Ho{\l}yst, Marek Litniewski, and Daniel Jakubczyk.
\newblock A molecular dynamics test of the hertz-knudsen equation for
  evaporating liquids.
\newblock {\em Soft Matter}, 11:7201, 2015.

\bibitem{Lebedev}
N.~N. Lebedev.
\newblock {\em Special Functions and Their Application}.
\newblock Prentice-Hall, Englewood Cliffs, New Jersey,, 1965.

\bibitem{Picknett1977}
R.G Picknett and R~Bexon.
\newblock The evaporation of sessile or pendant drops in still air.
\newblock {\em J. Colloid Interface Sci.}, 61(2):336--350, sep 1977.

\bibitem{HuHua2002}
Hua Hu and Ronald~G. Larson.
\newblock Evaporation of a sessile droplet on a substrate.
\newblock {\em J. Phys. Chem. B}, 106:1334, 2002.

\bibitem{Routh2013}
Alexander~F. Routh.
\newblock Drying of thin colloidal films.
\newblock {\em Rep. Prog. Phys.}, 76:046603, 2013.

\bibitem{Ma2005}
Yue Ma, H.T. Davis, and L.E. Scriven.
\newblock Microstructure development in drying latex coatings.
\newblock {\em Prog. Org. Coat.}, 52:46--62, 2005.

\bibitem{Arai2012}
S.~Arai and Masao Doi.
\newblock Skin formation and bubble growth during drying process of polymer
  solution.
\newblock {\em Eur. Phys. J. E}, 35:57, 2012.

\bibitem{Kajiya2006}
Tadashi Kajiya, Eisuke Nishitani, Tatsuya Yamaue, and Masao Doi.
\newblock Piling-to-buckling transition in the drying process of polymer
  solution drop on substrate having a large contact angle.
\newblock {\em Phys. Rev. E}, 73:011601, 2006.

\bibitem{deGennes2002}
P.~G. de~Gennes.
\newblock Solvent evaporation of spin cast films: ``crust'' effect.
\newblock {\em Eur. Phys. J. E}, 7:31--34, 2002.

\bibitem{Pauchard2003}
L.~Pauchard and C.~Allain.
\newblock Buckling instability induced by polymer solution drying.
\newblock {\em Europhys. Lett.}, 62:897, 2003.

\bibitem{Tsapis2005}
N.~Tsapis, E.~R. Dufresne, S.~S. Sinha, C.~S. Riera, J.W. Hutchinson,
  L.~Mahadevan, and D.~A. Weitz.
\newblock Onset of buckling in drying droplets of colloidal suspensions.
\newblock {\em Phys. Rev. Lett.}, 94:018302, 2005.

\bibitem{Boulogne2013a}
F.~Boulogne, F.~Giorgiutti-Dauphin{\'{e}}, and L.~Pauchard.
\newblock The buckling and invagination process during consolidation of
  colloidal droplets.
\newblock {\em Soft Matter}, 9:750--757, 2013.

\bibitem{Sadek2013}
C{\'{e}}line Sadek, Herv{\'{e}} Tabuteau, Pierre Schuck, Yannick Fallourd,
  Nicolas Pradeau, C{\'{e}}cile~Le Floch-Fou{\'{e}}r{\'{e}}, and Romain
  Jeantet.
\newblock Shape, shell, and vacuole formation during the drying of a single
  concentrated whey protein droplet.
\newblock {\em Langmuir}, 29:15606--15613, 2013.

\bibitem{MengFanlong2014}
Fanlong Meng, Masao Doi, and Zhongcan Ouyang.
\newblock Cavitation in drying droplets of soft matter solutions.
\newblock {\em Phys. Rev. Lett.}, 113:098301, 2014.

\bibitem{MengFanlong2016a}
Fanlong Meng, Ling Luo, Masao Doi, and Zhongcan Ouyang.
\newblock Solute based lagrangian scheme in modeling the drying process of soft
  matter solutions.
\newblock {\em Eur. Phys. J. E}, 39:22, 2016.

\bibitem{Masters}
K.~Masters.
\newblock {\em Spray Drying -- An Introduction to Principles, Operational
  Practice, and Applications}.
\newblock Leonard Hill Books, London, 1972.

\bibitem{Okuzono2006}
Tohru Okuzono, {Kin'ya} Ozawa, and Masao Doi.
\newblock Simple model of skin formation caused by solvent evaporation in
  polymer solutions.
\newblock {\em Phys. Rev. Lett.}, 97:136103, 2006.

\bibitem{LuoLing2016}
Ling Luo, Fanlong Meng, Junying Zhang, and Masao Doi.
\newblock Skin formation in drying a film of soft matter solutions: Application
  of solute based lagrangian scheme.
\newblock {\em Chin. Phys. B}, 25:076801, 2016.

\bibitem{Lintingre2016}
E.~Lintingre, F.~Lequeux, L.~Talini, and N.~Tsapis.
\newblock Control of particle morphology in the spray drying of colloidal
  suspensions.
\newblock {\em Soft Matter}, 12:7435--7444, 2016.

\bibitem{ChengShengfeng2013}
Shengfeng Cheng and Gary~S. Grest.
\newblock Molecular dynamics simulations of evaporation-induced nanoparticle
  assembly.
\newblock {\em J. Chem. Phys.}, 138:064701, 2013.

\bibitem{Reinhart2017}
Wesley~F. Reinhart, Andrew~W. Long, Michael~P. Howard, Andrew~L. Ferguson, and
  Athanassios~Z. Panagiotopoulos.
\newblock Machine learning for autonomous crystal structure identification.
\newblock {\em Soft Matter}, 13:4733--4745, 2017.

\bibitem{SuBin2012}
Bin Su, Yuchen Wu, and Lei Jiang.
\newblock The art of aligning one-dimensional (1d) nanostructures.
\newblock {\em Chem. Soc. Rev.}, 41:7832, 2012.

\bibitem{SuBin2013}
Bin Su, Yuchen Wu, Yue Tang, Yi~Chen, Wenlong Cheng, and Lei Jiang.
\newblock Free-standing 1d assemblies of plasmonic nanoparticles.
\newblock {\em Adv. Mater.}, 25:3968--3972, 2013.

\bibitem{msc_Crest}
J{\'e}r{\^o}me Crest.
\newblock Formation of microfibers and nanofibers by capillary-driven thinning
  of drying viscoelastic filaments.
\newblock Master's thesis, Massachusetts Institute of Technology, 2009.

\bibitem{Wooh2015}
Sanghyuk Wooh, Hannah Huesmann, Muhammad~Nawaz Tahir, Maxime Paven, Kristina
  Wichmann, Doris Vollmer, Wolfgang Tremel, Periklis Papadopoulos, and
  Hans-J\"{u}rgen Butt.
\newblock Synthesis of mesoporous supraparticles on superamphiphobic surfaces.
\newblock {\em Adv. Mater.}, 27:7338--7343, 2015.

\bibitem{Sekido2017}
Takafumi Sekido, Sanghyuk Wooh, Regina Fuchs, Michael Kappl, Yoshinobu
  Nakamura, Hans-J\"{u}rgen Butt, and Syuji Fujii.
\newblock Controlling the structure of supraballs by {pH}-responsive particle
  assembly.
\newblock {\em Langmuir}, 33:1995--2002, 2017.

\bibitem{Kumar2017}
Sanat~K. Kumar, Brian~C. Benicewicz, Richard~A. Vaia, and Karen~I. Winey.
\newblock 50th anniversary perspective: Are polymer nanocomposites practical
  for applications?
\newblock {\em Macromolecules}, 50:714--731, 2017.

\bibitem{Trueman2012}
R.~E. Trueman, E.~Lago Domingues, S.~N. Emmett, M.~W. Murray, and A.~F. Routh.
\newblock Auto-stratification in drying colloidal dispersions: A diffusive
  model.
\newblock {\em J. Colloid Interface Sci.}, 377:207, 2012.

\bibitem{Trueman2012a}
R.~E. Trueman, E.~Lago Domingues, S.~N. Emmett, M.~W. Murray, J.~L. Keddie, and
  A.~F. Routh.
\newblock Autostratification in drying colloidal dispersions: Experimental
  investigations.
\newblock {\em Langmuir}, 28:3420--3428, 2012.

\bibitem{Fortini2016}
Andrea Fortini, Ignacio Mart{\'i}n-Fabiani, Jennifer~Lesage De~La~Haye,
  Pierre-Yves Dugas, Muriel Lansalot, Franck {D'Agosto}, Elodie Bourgeat-Lami,
  Joseph~L. Keddie, and Richard~P. Sear.
\newblock Dynamic stratification in drying films of colloidal mixtures.
\newblock {\em Phys. Rev. Lett.}, 116:118301, 2016.

\bibitem{Howard2017}
Michael~P. Howard, Arash Nikoubashman, and Athanassios~Z. Panagiotopoulos.
\newblock Stratification dynamics in drying colloidal mixtures.
\newblock {\em Langmuir}, 33:3685--3693, 2017.

\bibitem{Zhou2017}
Jiajia Zhou, Ying Jiang, and Masao Doi.
\newblock Cross interaction drives stratification in drying film of binary
  colloidal mixtures.
\newblock {\em Phys. Rev. Lett.}, 118:108002, 2017.

\bibitem{Routh2004}
Alexander~F Routh and William~B Zimmerman.
\newblock Distribution of particles during solvent evaporation from films.
\newblock {\em Chem. Eng. Sci.}, 59:2961, 2004.

\bibitem{Kim2016b}
Sunhyung Kim, Kyu Hyun, Bernd Struth, Kyung~Hyun Ahn, and Christian Clasen.
\newblock Structural development of nanoparticle dispersion during drying in
  polymer nanocomposite films.
\newblock {\em Macromolecules}, 49:9068--9079, 2016.

\bibitem{ChengShengfeng2016}
Shengfeng Cheng and Gary~S. Grest.
\newblock Dispersing nanoparticles in a polymer film via solvent evaporation.
\newblock {\em {ACS} Macro Lett.}, 5:694, 2016.

\bibitem{Martin-Fabiani2016}
Ignacio Mart{\'{\i}}n-Fabiani, Andrea Fortini, Jennifer~Lesage de~la Haye,
  Ming~Liang Koh, Spencer~E. Taylor, Elodie Bourgeat-Lami, Muriel Lansalot,
  Franck D'Agosto, Richard~P. Sear, and Joseph~L. Keddie.
\newblock {pH}-switchable stratification of colloidal coatings: Surfaces
  {\textquotedblleft}on demand{\textquotedblright}.
\newblock {\em {ACS} Appl. Mater. Interfaces}, 8:34755--34761, 2016.

\bibitem{Atmuri2012}
Anand~K. Atmuri, Surita~R. Bhatia, and Alexander~F. Routh.
\newblock Autostratification in drying colloidal dispersions: Effect of
  particle interactions.
\newblock {\em Langmuir}, 28:2652, 2012.

\end{thebibliography}

\newpage

\end{document}